\documentclass[conference]{IEEEtran}
\IEEEoverridecommandlockouts
\usepackage{cite}
\usepackage{amsmath,amssymb,amsfonts,mathtools}
\usepackage{algorithm}
\usepackage[noend]{algpseudocode}
\usepackage{graphicx}
\usepackage{textcomp}
\usepackage{xcolor}
\def\BibTeX{{\rm B\kern-.05em{\sc i\kern-.025em b}\kern-.08em
    T\kern-.1667em\lower.7ex\hbox{E}\kern-.125emX}}

\newtheorem{defn}{Definition}

\DeclareMathOperator*{\argmin}{arg\,min}
\DeclarePairedDelimiter\abs{\lvert}{\rvert}%
\DeclarePairedDelimiter\norm{\lVert}{\rVert}%

\begin{document}

\title{Improving the Scalability of a Prosumer Cooperative Game with K-Means Clustering\\
\thanks{This work was supported in part by the Engineering and 
Physical Sciences Research Council under Grants EP/N03466X/1 and 
EP/S000887/1, and in part by the Oxford Martin Programme on Integrating 
Renewable Energy.}
\thanks{\textcopyright 2019 IEEE.  Personal use of this material is permitted.  Permission from IEEE must be obtained for all other uses, in any current or future media, including reprinting/republishing this material for advertising or promotional purposes, creating new collective works, for resale or redistribution to servers or lists, or reuse of any copyrighted component of this work in other works. }
\thanks{Digital Object Identifier 10.1109/PTC.2019.8810558.}
}

\author{\IEEEauthorblockN{Liyang Han, Thomas Morstyn, Constance Crozier, Malcolm McCulloch}
\IEEEauthorblockA{Department of Engineering Science \\
University of Oxford\\
Oxford, United Kingdom \\
\{Liyang.Han, Thomas.Morstyn, Constance.Crozier, Malcolm.McCulloch\}@eng.ox.ac.uk}
}

\maketitle

\begin{abstract}
Among the various market structures under peer-to-peer energy sharing, one model based on cooperative game theory provides clear incentives for prosumers to collaboratively schedule their energy resources. The computational complexity of this model, however, increases exponentially with the number of participants. To address this issue, this paper proposes the application of K-means clustering to the energy profiles following the grand coalition optimization. The cooperative model is run with the ``clustered players'' to compute their payoff allocations, which are then further distributed among the prosumers within each cluster. Case studies show that the proposed method can significantly improve the scalability of the cooperative scheme while maintaining a high level of financial incentives for the prosumers.
\end{abstract}


\section{Introduction}

K-means clustering is investigated in this paper as a means to address the computational issues embedded in the cooperative game theoretical model proposed in \cite{LHan2019}, which provides a framework to financially reward efficient collaboration of distributed energy resources (DER). 

As a result of the fast increase in DER and recent development in smart grid technologies, peer-to-peer (P2P) energy sharing or trading is widely proposed as a market mechanism \cite{Parag2016} to engage prosumers, proactive-consumers with distributed energy resources that actively control their energy behaviors.  


DER control strategies that incorporate optimal scheduling of energy storage (ES) have been heavily researched in microgrid applications where all of the participants align their interests with the microgrid \cite{Morstyn2017}. However, it is not necessarily the case that all prosumers can achieve their lowest energy costs individually while minimizing the joint energy cost. Therefore, it is important to study the strategic interactions among prosumers when designing local energy trading and sharing mechanisms.

Game theory is widely used in recent literature to link prosumers' energy behavior to their economic outcome. One typical example is the usage of non-cooperative game theory in energy sharing among prosumers, who strategically schedule their DER according to the dynamic dual prices \cite{LJia2016}. However, this scheme relies on the assumption that no prosumers have market power as dynamic dual prices can be interpreted as competitive energy prices, introducing potential instability to the practice \cite{ThomasMorstyn2018}. Besides, the allocation of the financial benefit to each prosumer may be unclear or suboptimal on the individual level \cite{LHan2019}. 

Cooperative game theory, on the other hand, details a clear profit allocation method that rewards collaboration, and analyzes the fairness of profit allocation from each individual player's perspective  \cite{Saad2012}. An energy sharing model that incorporates ES optimization was developed using cooperative game theory in \cite{LHan2019}, with the proof that the profit can be allocated in a way that ensures a satisfactory economic outcome for all prosumers in this P2P energy sharing scheme. However, the model's computational intensity remains a hindering factor due to the large number of linear optimization problems ($o(2^N)$ where $N$ is the number of players) required.

This paper aims to address this computational challenge. Sampling algorithms have been developed for cases where the Shapley value \cite{Shapley1971} is selected as the payoff allocation method \cite{Castro2009}. However, the Shapley value for this cooperative P2P energy sharing scheme is proven in \cite{LHan2019} to be sometimes non-stabilizing, meaning that some prosumers may have an incentive to leave the grand coalition to form smaller coalitions for higher profits, leading to inefficient utilization of their collective energy storage capacity. The nucleolus, despite being stabilizing for this application, is calculated by iteratively minimizing all possible coalitions' excesses \cite{Sankaran1991}, introducing another intractable step into the model.

Under the cooperative theoretical scheme, the profit allocation is directly linked to each player's contribution to the coalitions, which is measured by how much a player's own energy behavior can offset the inefficiencies in the coalitional energy usage. Therefore, we consider grouping customers with similar load patterns into joint players in order to limit the number of possible coalitions, thus reducing the required number of linear optimization problems. Energy profile clustering has been extensively used to study the customer load patterns \cite{1626400, Petitjean2011, Chen2017}. One common objective among these works is using typical load patterns to inform the setting of tariff structures, but they have not considered the flexibility of storage, or analyzed the added cost savings of a certain load profile to a cooperative group, which is the determining factor in a prosumer's profit allocation. 

We propose, in this paper, the novel concept of ``clustered players'', who are grouped by applying K-means to their load profiles under the cooperative energy management scenario. We run the cooperative game model with just the ``clustered players'' instead of all the participating prosumers, significantly reducing the number of linear problems. The payoff allocation for the ``clustered players'' can then be further distributed among their member prosumers. Finally, We use case studies to demonstrate the significant improvement in the model's scalability and a similar level of financial incentives for the prosumers using the proposed method. It is thus a good benchmark for future development of customized clustering methods for the cooperative P2P energy sharing scheme.

\section{Cooperative Game Formulation}

In order to formulate the collaboration of different prosumers as a cooperative game, we need to answer three key questions: 1) how do prosumers collaborate? 2) how do we quantify the value of this collaboration? and 3) how do we allocate the benefit gained through collaboration to each individual player? The following three subsections provide detailed discussions on these questions.

\subsection{Cooperative Energy Management}

Energy cost savings are achieved through scheduling flexible energy resources, usually energy storage (ES) systems, to benefit from time-of-use (ToU) energy pricing and dual energy pricing, in which retail contracts pay prosumers less for excess generation than they charge for energy usage \cite{Zhou2018}.

Under the assumption that the retail supplier offers a lower export tariff (e.g. feed-in tariff) than it charges for buying electricity, prosumers can save money by storing the excess PV generation in the ES system and using it when the generation is less than they need. The concept of an \emph{energy coalition} was proposed in \cite{8443054}, where a group of prosumers collaboratively operate their ES systems to minimize the total group energy cost. Let a group of \(N\) prosumers form \emph{grand coalition} \(\mathcal{N}\) indexed by \(\mathnormal{i} \in \mathcal{N} := \{1,2,...,N\}\). A \emph{coalition} is any subset \(\mathcal{S} \subseteq \mathcal{N}\). Assuming  \(R\) timesteps (\(t = 1, 2,..., R\)) with a time interval of \(\Delta t\), the total energy cost for a coalition $\mathcal{S}$ can be written as a function of all its members' ES operations:
\begin{equation*} \label{sc_ec}
F_{\mathcal{S}}(\mathbf{b}) = \sum_{t=1}^{R} \Big\{p^{b}_{t} \sum_{i \in \mathcal{S}} [q_{it} + b_{it}]^+ + p^{s}_{t} \sum_{i \in \mathcal{S}} [q_{it} + b_{it}]^- \Big\}
\end{equation*}
where \([x]^{+(-)} = \max (\min) \{x,0\}\). $q_{it}$, in $[\mathit{kWh}]$, denotes prosumer \(i\)'s net energy consumption (positive) or generation (negative) without ES at time \(t\), $b_{it}$, in $[\mathit{kWh}]$, denotes prosumer \(i\)'s ES system's charge (positive) or discharge (negative) energy variables at time \(t\), and $p^{b}_{t}$ and $p^{s}_{t}$, in $[\pounds/\mathit{kWh}]$, denote the electricity import and export prices at time \(t\).

The \textit{coalitional energy cost} for \(\mathcal{S}\) is then defined as the lowest total energy cost achievable by optimizing the operation of all the ES systems within \(\mathcal{S}\) at the same time:
\begin{align}
C(\mathcal{S}) = \ & \min_{\mathbf{b}}  {F_\mathcal{S}(\mathbf{b})} \label{sc_min_ec} \\
s.t. \ \ & \underline{b}_{i} \leq b_{it} \leq \overline{b}_{i}, \ \ \forall i \in \mathcal{S}, \forall t \in [1, R] \label{p_cnstr} \\
& 0 \leq e_{i} SoC^{0}_i + \sum_{t=1}^k ([b_{it}]^{+} \eta_{i}^{in} + [b_{it}]^{-} / \eta_{i}^{out})  \leq e_{i}, \nonumber \\
& \qquad \qquad \qquad \qquad \qquad \quad \forall i \in \mathcal{S}, \forall r \in [1, R] \label{e_cnstr} \\
& \sum_{t=1}^R ([b_{it}]^{+} \eta_{i}^{in} + [b_{it}]^{-} / \eta_{i}^{out}) = 0, \ \ \forall i \in \mathcal{S} \label{c_cnstr}
\end{align}
where each prosumer \(i\)'s ES system has an energy capacity of \(e_{i}\), in \([\mathit{kWh}]\), charge limit $\overline{b}_{i}$ and discharge limit $\underline{b}_{i}$ over \(\Delta t\), in \([\mathit{kWh}]\), charge efficiency \(\eta_{i}^{in}\) and discharge efficiency \(\eta_{i}^{out} \), and initial state of charge \(SoC^{0}_i\). 

The ES power constraint, energy constraint, and cycle constraint are expressed respectively in (\ref{p_cnstr}), (\ref{e_cnstr}), and (\ref{c_cnstr}). The piecewise components of this optimization problem can be rewritten in a linearized format detailed in  \cite{8443054}.

The \textit{coalitional energy cost} through solving (\ref{sc_min_ec}) then serves as a basis to evaluate the value of each energy coalition.

\subsection{Value of Energy Coalitions}

For each \textit{energy coalition} \(\mathcal{S}\) we assign a \textit{value function} \(v(\mathcal{S})\), defined as the energy cost savings by forming \(\mathcal{S}\), in other words, the difference between the sum of the energy costs of each prosumer in \(\mathcal{S}\) with ES systems scheduled individually, and the \textit{coalitional energy cost} of \(\mathcal{S}\) with all the ES systems scheduled collaboratively: 
\[v(\mathcal{S}) = \sum_{i \in \mathcal{S}} {C({\{i\}})} - C({\mathcal{S}}) \]

The pair \( (\mathcal{N}, v) \) defines our prosumer cooperative game, and the \textit{value} of the \textit{grand coalition} \(v(\mathcal{N})\) denotes the total amount of payoffs we can award to the prosumers.

\subsection{Payoff Allocation}

Having calculated the total amount of payoffs available, the next step is to determine how to allocate it to each prosumer.

\begin{defn} [Imputation]
We use vector \(\mathbf{x}\) as the \textit{payoff allocation}, and its entry \(x_i\) represents the payment to prosumer \(i \in \mathcal{N}\). \(\mathbf{x}\) is said to be an imputation if it meets both the \textit{Efficiency} and \textit{Individual Rationality} criteria:
\end{defn}
\begin{enumerate}
\item \emph{Efficiency}: \(\sum_{i \in \mathcal{N}} {x_i} = v(\mathcal{N})\).
\item \emph{Individual Rationality}: \(x_i \geq v(\{i\}), \forall i \in \mathcal{N}\).
\end{enumerate}
The \emph{efficiency} criterion guarantees the sum of all \emph{payoff allocations} equals the \textit{grand coalition}'s energy cost savings, and the \emph{individual rationality} criterion requires all the prosumers are better off cooperating in the \textit{grand coalition}.

However, an imputation does not guarantee everyone being \textit{satisfied} in the \textit{grand coalition}, as some  players may be able to achieve higher payoffs by forming smaller \textit{energy coalitions}.
\begin{defn} [Excess]
We measure an \textit{energy coalition}'s \textit{dissatisfaction} with respect to the imputation \(\mathbf{x}\) by its excess defined as $\varepsilon(\mathcal{\mathbf{x}, S}) = v(\mathcal{S}) - \sum_{ i \in \mathcal{S}} {x_i} $.
\end{defn}

If \(\varepsilon(\mathcal{\mathbf{x}, S}) >0 \), then \(\mathcal{S}\) is better off on its own and can offer higher payoffs to its members than the \textit{grand coalition}. 

The nucleolus $\mathbf{u}$, the imputation with lexicographically minimal excesses that minimizes the dissatisfaction of the players \cite{Baeyens2013}, is proven in \cite{LHan2019} to be \emph{stabilizing} for our prosumer cooperative game, ensuring $\varepsilon(\mathcal{\mathbf{u}, S}) \leq 0, \forall \mathcal{S} \subseteq \mathcal{N} $.

The significance of this cooperative game theoretic approach is that it offers a way to encourage DER collaboration while ensuring that all players are financially incentivized to remain in the \emph{energy grand coalitions}. A key limitation, however, is the computational complexity which increases exponentially with the number of prosumers. The following section explores the potential of clustering in addressing this problem.

\section{Prosumer Clustering}

The scalability of this cooperative game model is mainly limited by two groups of computation steps: 1) cost minimization for all \(2^N\) coalitions to calculate the \emph{coalitional energy costs}, and 2) excess minimization for all \(2^N\) coalitions to calculate the \emph{nucleolus}. All optimization problems in each step are linear and tractable, but it is the sheer number of linear problems in each step that proves intractable as it increases exponentially with the number of players. 

We identified clustering as a means to reduce the number of coalitions, hence the number of linear optimization problems required. Clustering seeks to group similarities and separate differences among its subjects, which we define as prosumer load patterns in our cooperative game. This matches well with our objective to allocate payoffs to the prosumers based on the contribution they make to each coalition, which is measured by their capacity to \emph{offset} the net consumption or generation of the rest of the coalition.

\subsection{K-Means Clustering}

Various clustering algorithms have been used and compared in identifying load patterns, most commonly to inform the setting of tariff structures. However, these are not necessarily suitable for partitioning prosumers in a cooperative energy management scheme. As a benchmark, this paper incorporates a simple K-means clustering algorithm in the prosumer cooperative game as an attempt to reduce the full game's computational complexity.

In implementing K-means clustering, some initial seed selections can lead to local optimal solutions \cite{Meila2006}. In this paper, we choose 24 hours as the model timespan and half-hourly energy consumption values as the clustering features. Due to the low number of features (i.e. 48 features for one day) and the small sample size (i.e. \(\leq 200\) prosumers), we could simply apply a random initialization and run the K-means model 1000 times without significantly impacting the computational time. An important benefit of this approach is that we could compare all the K-means results and select the final clusters based on additional criteria. As the selection of K is limited by the computational complexity, we would like the prosumers as evenly represented in the clusters as possible. Therefore, we choose the total Euclidean distance of each K-means model's result as a metric, and identify a set of clustering results that have the lowest total Euclidean distances by setting a selection upper bound as 1\% above the lowest total Euclidean distance of all runs. All the clustering results that fall within the relaxed range of total Euclidean distance can then be compared against each other, and the one with the most even distribution of prosumers is selected. A detailed description of implementing this K-means clustering technique is shown in Algorithm \ref{opt_centroids}.

\begin{algorithm}
\caption{Cluster Assignment (input: profiles, k)}\label{opt_centroids}
\begin{algorithmic}[0]
\Function{Kmeans}{$\mathbf{m}, k$}
\State {(cluster prof.) $\mathbf{p} \gets \argmin_\mathbf{P}\sum_{j=1}^k\sum_{\mathbf{m} \in P_j} \norm{\mathbf{m}-\mathbf{p}_j}^2$}
\State {(total Euclidean dist.) $D \gets \sum_{j=1}^k\sum_{\mathbf{m} \in P_j} \norm{\mathbf{m}-\mathbf{p}_j}$}
\State {(cluster sizes) $\mathbf{P}^\ast: P^\ast_j \gets \text{count of }\textbf{m} \in P_j, \forall j \in [1,k]$}
\State {(cluster assignment) $\mathbf{G}: G_i \gets j \text{ if } m_i \in P_j, \forall i$}
\State \Return {$\mathbf{p}, D, \mathbf{P}^\ast, \mathbf{G}$}
\EndFunction
\hrulefill
\State (total K-means runs) $\textit{runmax} \gets 1000$
\State (prosumer load profiles) $\textbf{prof}$
\State (total Euclidean dist. selection relaxation) $\textit{eurelax} \gets 1\%$
\For {$\textit{r} \textbf{ in } [1, \textit{runmax}]$}
\State {$\mathbf{c}[r], \textit{EU}[r], \mathbf{C}^\ast[r], \mathbf{M}[r] \gets$ \Call{Kmeans}{$\textbf{prof}, k$}}
\EndFor
\State {$\textit{eumin}  \gets min{(\textit{EU})}, \textit{minr} \gets \argmin{(\textit{EU})}$}
\State {$\textbf{c}_a \gets \textbf{c}[\textit{minr}], \textbf{C}_a^\ast \gets \textbf{C}^\ast[\textit{minr}], \textbf{M}_a \gets \mathbf{M}[\textit{minr}]$}
\For {$\textit{r} \textbf{ in } [1, \textit{runmax}]$}
\If {$\textit{EU}[r] \leq \textit{eumin}*(1+\textit{eurelax})$}
\If {$min(\mathbf{C}^\ast[r]) > min(\textbf{C}_a^\ast)$}
\State {$\textbf{c}_a \gets \textbf{c}[r], \textbf{C}_a^\ast \gets \textbf{C}^\ast[r], \textbf{M}_a \gets \mathbf{M}[r]$}
\ElsIf {$min(\mathbf{C}^\ast[r]) = min(\mathbf{C}_a^\ast) $}
\If {$max(\mathbf{C}^\ast[r]) < max(\mathbf{C}_a^\ast)$}
\State {$\textbf{c}_a \gets \textbf{c}[r], \textbf{C}_a^\ast \gets \textbf{C}^\ast[r], \textbf{M}_a \gets \mathbf{M}[r]$}
\EndIf
\EndIf
\EndIf
\EndFor
\State \Return $\textbf{C}_a^\ast, \textbf{M}_a$

\end{algorithmic}
\end{algorithm}

\begin{figure}
\begin{center}
\includegraphics[width=8.4cm]{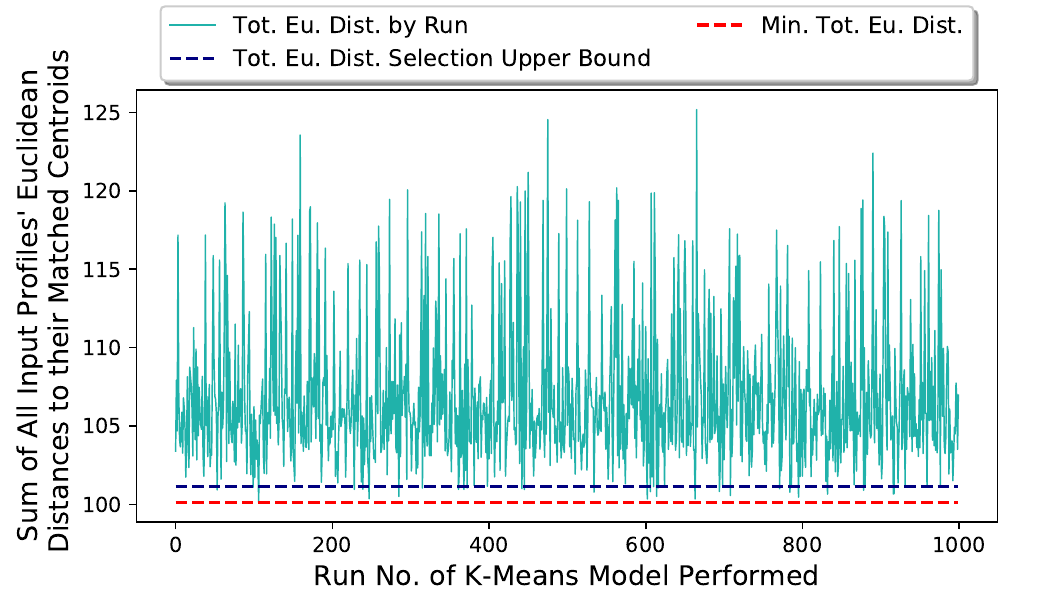}    
\caption{Total Euclidean distances for 1000 K-means (k=8) clustering runs on 50 energy profiles}
\label{fig:euclidean_dist}
\end{center}
\vspace*{-0.4cm}
\end{figure}

Fig.~\ref{fig:euclidean_dist} shows an example of the Euclidean distance variation as a result of clustering 50 prosumer load profiles. It can be seen than at least 20 out of 1000 clustering runs produce a total Euclidean distance within the allowed range. To have this adjusted K-means algorithm provides the flexibility to apply additional screening methods to the selection of prosumer clusters.

\subsection{Cooperative Game with Clustered Players} \label{subsec:cl_coop}

After the clustering method is finalized, we still need to decide which set of energy profiles to apply clustering to. Because the prosumer cooperative game focuses on the contributions of each player to the coalitions, we select the net load profiles under the cooperative \emph{grand coalition} scenario as the clustering subjects. This means we need to run the cooperative energy management model for the \emph{grand coalition} with all of the $N$ players first before clustering can be applied.

The purpose of clustering in this paper is to reduce the number of modeled players in the cooperative game model. Therefore, we define each cluster of prosumers as a \emph{clustered player} $cl_j$, and the model becomes a $K$-player cooperative game, where $K$ is the input number of clusters for K-means. We then have \(cl_j \in cl_\mathcal{K} := \{cl_1,cl_2,...,cl_K\}\). Any coalition of clustered players is defined as $cl_\mathcal{U} \subseteq cl_\mathcal{K} $. While running the cooperative game model for the clustered players, we assign to each clustered player the sum of all its member prosumers' original consumption and generation profiles as its own consumption and generation, and all of its member prosumers' ES systems as its own ES capacity. 

In other words, we follow the following steps to implement this cooperative game model with clustered players.
\begin{enumerate}
\item \emph{Load processing}: we gather all the model inputs, including the number of prosumers $N$, their net load profiles $\mathbf{q}: q_{it}$, where $i \in [1, N], t \in [1, R]$, and their ES system constraints. We calculate the \emph{coalitional energy cost} for the \emph{grand coalition} $C(\mathcal{N})$ following (\ref{sc_min_ec}), and record the optimized ES operation as $\mathbf{b}^\ast: b^\ast_{it}$. 

\item \emph{Prosumer clustering}: we construct each prosumer's load profile as a result of the cooperative energy management in the \emph{grand coalition}: $\mathbf{l}^\ast: l^\ast_{it} =  q_{it} + b^\ast_{it}$. The K-means clustering model detailed in Algorithm \ref{opt_centroids} is then run with $\mathbf{l}^\ast$ as inputs, along with a chosen $K$. We obtain the clustering assignment $\mathbf{a}: a_i = j \mid i \in cl_j, \forall i \in \mathcal{N}$. Therefore, for any player \(i\), the cluster it belongs to can be notated as \(cl_{a_i}\).
\begin{figure*}
\begin{center}
\includegraphics[width=\textwidth]{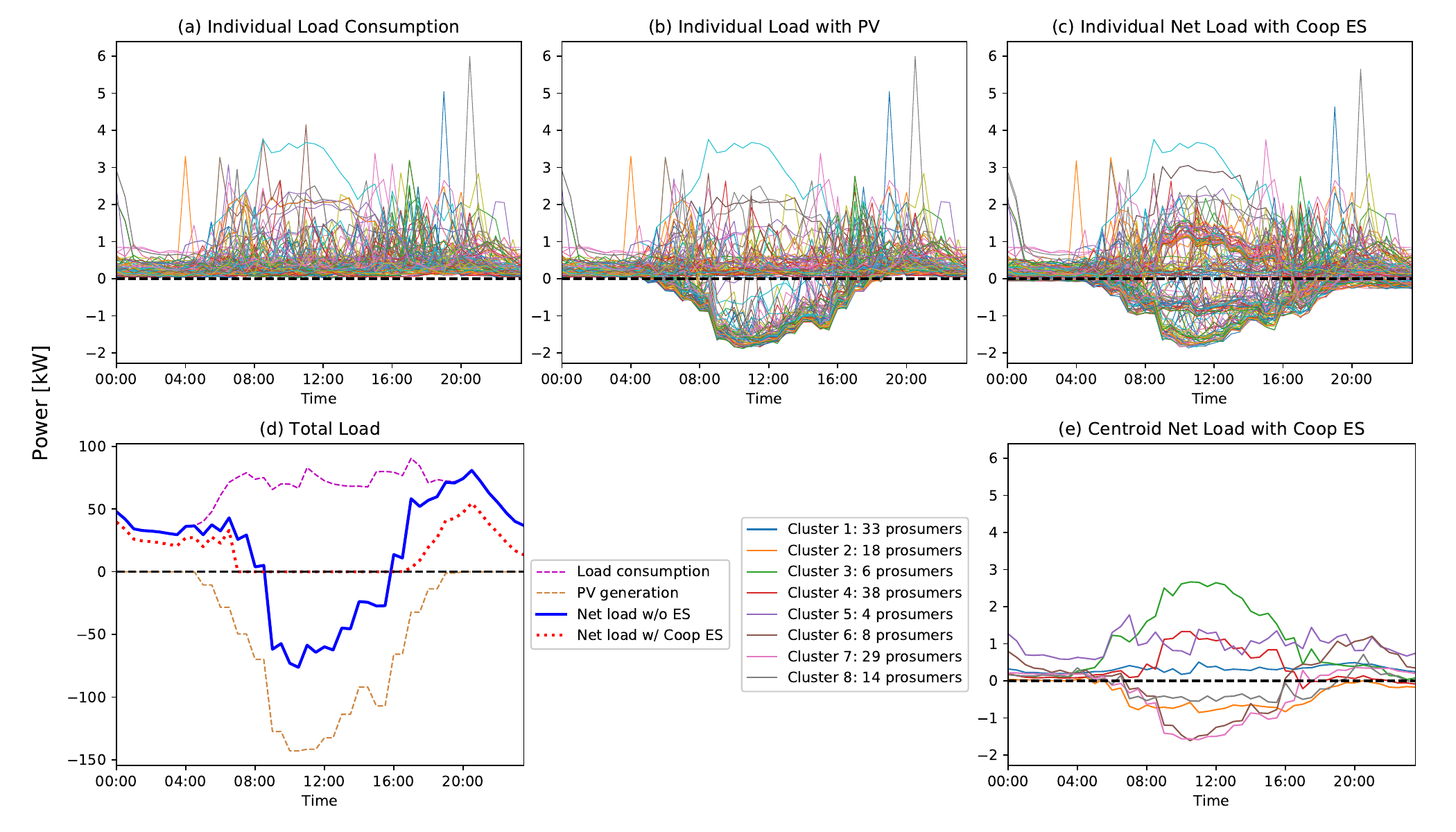}
\caption{Plots of 150 prosumers' (a) load consumption alone, (b) net load with PV without ES operation, (c) net load with PV and ES optimized for the grand coalition, (d) joint loads with and without cooperative ES operation, (e) 8 centroid net load profiles with ES optimized for the grand coalition.}
\label{fig:clustering_eg}
\end{center}
\vspace*{-0.2cm}
\end{figure*}
\item \emph{Clustered player cooperative game formulation}: in order to preserve as much of the original inputs as possible, we apply all the prosumers' load and ES inputs directly to the clustered player coalitions. Therefore, for any coalition of clustered players $cl_\mathcal{U} \subseteq cl_\mathcal{K}$, the total energy cost is defined as
\begin{multline*}
F_{cl_\mathcal{U}}(\mathbf{b}) = \sum_{t=1}^{R} \Big\{p^{b}_{t} \sum_{cl_{a_i} \in cl_\mathcal{U}} [q_{it} + b_{it}]^+ \\
+ p^{s}_{t} \sum_{cl_{a_i} \in cl_\mathcal{U}} [q_{it} + b_{it}]^- \Big\}
\end{multline*}
This way we convert all the clustered player inputs to the original prosumers' inputs, so we are able to use (\ref{sc_min_ec}) to compute the \emph{coalitional energy costs} for all the clustered player coalitions $C(cl_\mathcal{U})$. Using a similar conversion, the value of clustered player coalitions can be calculated as
\begin{equation*} 
v(cl_\mathcal{U}) = \sum_{cl_{a_i}  \in cl_\mathcal{U}} {C({\{i\}})} - C(cl_\mathcal{U})
\end{equation*}
The nucleolus of this clustered player cooperative game $\mathbf{u}: u_{cl_j}, \forall cl_j \in cl_\mathcal{K}$ can then be computed by iteratively minimizing the excess of the clustered player coalitions \cite{8443054}.
\item \emph{De-clustering}: the nucleolus computed for the clustered players needs to be distributed to their member prosumers. Different from the full prosumer cooperative model where the value for each coalition of a single prosumer $v(\{i\}) = 0, \forall i \in \mathcal{N}$, the in-cluster cooperation sometimes results in $v(\{cl_j\})>0$. This can be considered additional savings for each clustered player that also needs to be distributed to its member prosumers. Although advanced allocation methods can be developed, to set a benchmark, here we distribute the energy savings to each prosumer within a cluster simply in proportion to the absolute value of each individual's energy cost without cooperation:
\begin{equation*}
\mathbf{x}: x_i = [u_{cl_{a_i}} + v(\{cl_{a_i}\})] (\abs{C({\{i\}})}/ \sum_{i \in cl_j} \abs{C({\{i\}})} )
\end{equation*}
\end{enumerate}

K-means clustering reduces the number of coalitions of the cooperative game from $2^N$ to $2^K$, where $K \leq N$. Considering $K$ can theoretically take on any value we choose, we now have full control of the model's computation complexity. Intuitively, the lower $K$ is, the more information the model loses, impacting the model's accuracy. An example is discussed in the case studies. 

\section{Case Studies}

In this section, we incorporate clustering in our prosumer cooperative game for two case studies. We use domestic load data measured in the Customer-Led Network Revolution trials\footnote{http://www.networkrevolution.co.uk/resources/project-library/}. We select a time frame of 24 hours starting from the midnight of a sunny summer day in July. The PV generation data is simulated in PVWatts \footnote{http://pvwatts.nrel.gov/pvwatts.php} using a 4kW, fixed 20 degree tilt residential system, and the London Gatwick solar data. The ES model has an \textit{energy capacity} of 7 kWh, a maximum charge power of 3.5 kW, a maximum discharge power of 3.2 kW, both charge and discharge efficiencies of 95\%, an \textit{initial state of charge} of 50\%, and a \textit{state of charge} range of 20-95\%. We assume PV and ES are adopted by 50\% of the prosumers respectively, and both ownerships are randomly assigned independently of each other. In other words, each prosumer can have a PV system, or an ES system, or both, or neither. The energy import price follows a UK Economy 7 residential rate structure: \pounds 0.072/kWh for midnight--7am, and \pounds 0.1681/kWh for 7am--midnight\footnote{https://www.gov.uk/government/statistical-data-sets/annual-domestic-energy-price-statistics}, and the energy export price is the UK feed-in tariff\footnote{https://www.gov.uk/feed-in-tariffs/overview} fixed at \pounds 0.0485/kWh. 

\subsection{Model with a Large Number of Prosumers} \label{case_largeNo}

We run the clustering model for large numbers of players ($14 \leq N \leq 200$) to test its effectiveness in grouping prosumers based on their DER mix and reducing computation time. Fig.~\ref{fig:clustering_eg} shows an example of clustering 150 prosumers into 8 groups following the steps detailed in Section \ref{subsec:cl_coop}. Increasing the number of clusters can improve the accuracy of the nucleolus estimation, but it will also result in an increase in the computation time. In order to maintain a balance between the two, we chose 8 clusters, double the number of the types of energy resource mix (i.e. no PV or ES, only PV, only ES, and both PV and ES).

\begin{table}[t]
\vspace*{-0.1cm}
\begin{center}
\caption{Clustering Results by DER}\label{tb:cl_der}
\vspace*{-0.25cm}
\begin{tabular}{ccccccccc}
\hline
\textbf{Clustered player} & \textbf{1} & \textbf{2} & \textbf{3} & \textbf{4} & \textbf{5} & \textbf{6} & \textbf{7} & \textbf{8} \\\hline
No. of prosumers & 10 & 28 & 7 & 9 & 22 & 6 & 30 & 38 \\
No. of PV units & 10 & 28 & 0 & 9 & 22 & 0 & 4 & 2 \\
No. of ES systems & 10 & 0 & 0 & 0 & 22 & 4 & 2 & 37 \\\hline
\end{tabular}
\end{center}
\vspace*{-0.25cm}
\end{table}

\begin{table}[t]
\begin{center}
\caption{Model Computation Time [s]}\label{tb:comp_t}
\vspace*{-0.14cm}
\begin{tabular}{cccccccc}
\hline
\textbf{No. prosumers} & \textbf{8} & \textbf{10} & \textbf{14} & \textbf{50} & \textbf{100} & \textbf{200} \\\hline
w/o clustering & 13 & 77 & 3E+4 & N/A & N/A & N/A \\
with \textbf{8} clusters & 13 & 24 & 54 & 6E+2 & 5E+3 & 3E+4 \\\hline
\end{tabular}
\end{center}
\vspace*{-0.25cm}
\end{table}

Table~\ref{tb:cl_der} shows the DER distribution in the clusters. It can be clearly observed that clustering effectively separated prosumers with different types of DER; Clustered Player 1, 5 consist of prosumers with both PV and ES, Clustered Player 2, 4 with only PV, Clustered Player 3, 7 mostly without PV or ES, Clustered Player 6, 8 mostly with only ES. Table~\ref{tb:comp_t} compares the computation time\footnote{Running on Apple iMac with a processor of 2.8 GHz Intel Core i5 and a memory module of 16 GB 1867 MHz DDR3} between the model with clustering and the model without clustering. It shows that the model without clustering requires about 10 hours to compute for a 14-prosumer game, whereas the model with clustering can solve a 200-prosumer game within the same amount of time.

\subsection{Full Model vs. Model with Clustering}

In this case study, we aim to verify the accuracy of a model with clustering by comparing its results to a full model. Due to the computational limitation of a full model, we can only conduct the comparison with a relatively low number of prosumers. Here we select 14 prosumers with exactly the same load and DER inputs for both models, and 5 clusters for the model with clustering. Again, 5 was chosen as the number of clusters to balance the accuracy and computational time.

\begin{figure}
\vspace*{-0.5cm}
\begin{center}
\includegraphics[width=8.4cm]{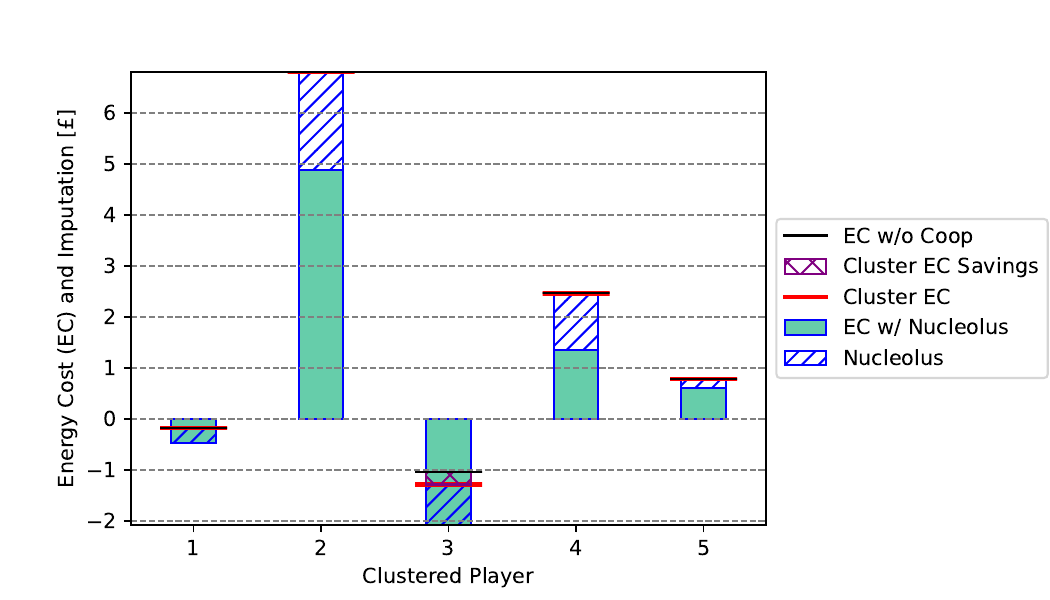}    
\caption{Energy costs (EC) and imputation for clustered players} 
\label{fig:cl_imputation}
\end{center}
\vspace*{-0.2cm}
\end{figure}

Fig.~\ref{fig:cl_imputation} displays the energy costs, cost savings from both cooperating within the clusters and the final imputation (i.e. nucleolus). It can be seen that all the clustered players are guaranteed savings, and the savings from cooperating within the clusters are significantly lower than the nucleolus, which further confirms that the clustering process tends to group prosumers that exhibit similar behaviors together. 

\begin{figure}
\begin{center}
\includegraphics[width=7.5cm]{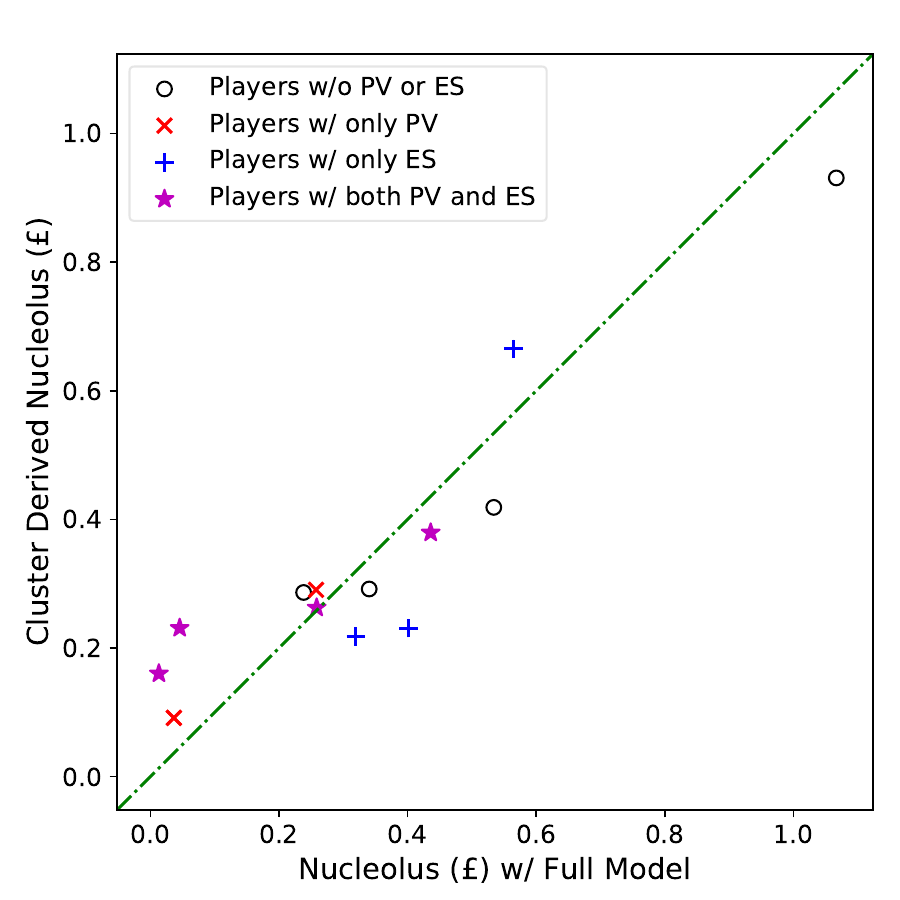}    
\caption{Nucleolus comparison (full model vs. model w/ clustering)} 
\label{fig:exc_comp}
\end{center}
\vspace*{-0.4cm}
\end{figure}

Fig.~\ref{fig:exc_comp} compares the two sets of nucleolus calculated in both models. Each marker represents a prosumer, and they all fall very close to the diagonal line regardless of the player DER mix, which means the payoff allocation computed from clustering is a good estimation of the nucleolus computed through a full model. The result discrepancies here are mainly due to the `lumping' effect of clustering, which is unable to capture the individual contributions of each player. More advanced de-clustering techniques can be developed for the re-distribution of payoff among each cluster to improve the estimation accuracy. 


\section{Conclusion}

To overcome the computational challenge in the cooperative P2P energy sharing scheme, we incorporated K-means clustering in the model. It was shown to effectively sort prosumers into groups that were representative of each individual's contribution to the cooperation. As a result, the computation time was significantly reduced and the model with clustering was able to solve for 200 players within the time a full model would take for 14 players. The payoff allocation derived from our benchmark clustering and de-clustering methods was shown to be very close to the nucleolus calculated in the full model. Considering prosumer's generation and consumption vary on a daily basis, it is unlikely for a player or coalition to be under-compensated consistently that would give it an incentive to become independent. Further analyses are needed to confirm this hypothesis. A natural extension of this work is to further customize the clustering and de-clustering techniques so that the modified cooperative game model can be applied to larger numbers of prosumers while maintaining the incentives for the participants at a similar level as the full model.


\bibliographystyle{myIEEEtran}
\bibliography{bib_2019_LH.bib} 

\begin{thebibliography}{10}
\providecommand{\url}[1]{#1}
\csname url@rmstyle\endcsname
\providecommand{\newblock}{\relax}
\providecommand{\bibinfo}[2]{#2}
\providecommand\BIBentrySTDinterwordspacing{\spaceskip=0pt\relax}
\providecommand\BIBentryALTinterwordstretchfactor{4}
\providecommand\BIBentryALTinterwordspacing{\spaceskip=\fontdimen2\font plus
\BIBentryALTinterwordstretchfactor\fontdimen3\font minus
  \fontdimen4\font\relax}
\providecommand\BIBforeignlanguage[2]{{%
\expandafter\ifx\csname l@#1\endcsname\relax
\typeout{** WARNING: IEEEtran.bst: No hyphenation pattern has been}%
\typeout{** loaded for the language `#1'. Using the pattern for}%
\typeout{** the default language instead.}%
\else
\language=\csname l@#1\endcsname
\fi
#2}}

\bibitem{LHan2019}
L.~{Han}, T.~{Morstyn}, and M.~{McCulloch}, ``Incentivizing prosumer coalitions
  with energy management using cooperative game theory,'' \emph{IEEE
  Transactions on Power Systems}, vol.~34, no.~1, pp. 303--313, Jan 2019.

\bibitem{Parag2016}
Y.~Parag and B.~K. Sovacool, ``{Electricity market design for the prosumer
  era},'' \emph{Nature Energy}, no. March, p. 16032, 2016.

\bibitem{Morstyn2017}
T.~Morstyn, B.~Hredzak, R.~P. Aguilera, and V.~G. Agelidis, ``Model predictive
  control for distributed microgrid battery energy storage systems,''
  \emph{IEEE Transactions on Control Systems Technology}, vol.~26, no.~3, pp.
  1107--1114, May 2018.

\bibitem{LJia2016}
L.~Jia and L.~Tong, ``Dynamic pricing and distributed energy management for
  demand response,'' \emph{IEEE Transactions on Smart Grid}, vol.~7, no.~2, pp.
  1128--1136, March 2016.

\bibitem{ThomasMorstyn2018}
T.~{Morstyn}, A.~{Teytelboym}, and M.~D. {Mcculloch}, ``Bilateral contract
  networks for peer-to-peer energy trading,'' \emph{IEEE Transactions on Smart
  Grid}, vol.~10, no.~2, pp. 2026--2035, March 2019.

\bibitem{Saad2012}
W.~Saad, Z.~Han, H.~V. Poor, and T.~Basar, ``Game-theoretic methods for the
  smart grid: An overview of microgrid systems, demand-side management, and
  smart grid communications,'' \emph{IEEE Signal Processing Magazine}, vol.~29,
  no.~5, pp. 86--105, Sept 2012.

\bibitem{Shapley1971}
L.~S. Shapley, ``{Cores of convex games},'' \emph{International Journal of Game
  Theory}, vol.~1, no.~1, pp. 11--26, 1971.

\bibitem{Castro2009}
J.~Castro, D.~G{\'{o}}mez, and J.~Tejada, ``{Polynomial calculation of the
  Shapley value based on sampling},'' \emph{Computers and Operations Research},
  vol.~36, no.~5, pp. 1726--1730, 2009.

\bibitem{Sankaran1991}
J.~K. Sankaran, ``{On finding the nucleolus of an n-person cooperative game},''
  \emph{International Journal of Game Theory}, vol.~19, no.~4, pp. 329--338,
  1991.

\bibitem{1626400}
G.~Chicco, R.~Napoli, and F.~Piglione, ``Comparisons among clustering
  techniques for electricity customer classification,'' \emph{IEEE Transactions
  on Power Systems}, vol.~21, no.~2, pp. 933--940, May 2006.

\bibitem{Petitjean2011}
F.~Petitjean, A.~Ketterlin, and P.~Gan{\c{c}}arski, ``{A global averaging
  method for dynamic time warping, with applications to clustering},''
  \emph{Pattern Recognition}, vol.~44, no.~3, pp. 678--693, 2011.

\bibitem{Chen2017}
T.~Chen, K.~Qian, A.~Mutanen, B.~Schuller, P.~Jarventausta, and W.~Su,
  ``{Classification of electricity customer groups towards individualized price
  scheme design},'' \emph{2017 North American Power Symposium, NAPS 2017},
  no.~1, pp. 4--7, 2017.

\bibitem{Zhou2018}
Y.~Zhou, J.~Wu, and C.~Long, ``{Evaluation of peer-to-peer energy sharing
  mechanisms based on a multiagent simulation framework},'' \emph{Applied
  Energy}, vol. 222, no. February, pp. 993--1022, 2018.

\bibitem{8443054}
L.~Han, T.~Morstyn, and M.~McCulloch, ``Constructing prosumer coalitions for
  energy cost savings using cooperative game theory,'' in \emph{2018 Power
  Systems Computation Conference (PSCC)}, June 2018, pp. 1--7.

\bibitem{Baeyens2013}
E.~Baeyens, E.~Y. Bitar, P.~P. Khargonekar, and K.~Poolla, ``{Coalitional
  aggregation of wind power},'' \emph{IEEE Transactions on Power Systems},
  vol.~28, no.~4, pp. 3774--3784, 2013.

\bibitem{Meila2006}
M.~Meil\u{a}, ``The uniqueness of a good optimum for k-means,'' in
  \emph{Proceedings of the 23rd International Conference on Machine Learning},
  ser. ICML '06.\hskip 1em plus 0.5em minus 0.4em\relax New York, NY, USA: ACM,
  2006, pp. 625--632.

\end{thebibliography}

\end{document}